\documentclass[final,5p,times,twocolumn]{elsarticle}

\usepackage{amssymb}
\usepackage{amsmath}
\usepackage{subcaption}
\usepackage{listings}
\usepackage{xcolor}
\usepackage{booktabs}
\usepackage{siunitx}
\usepackage{multirow, makecell}
\usepackage{tikz}
\usetikzlibrary{positioning}
\usepackage[acronyms]{glossaries}
\usepackage{hyperref}
\newacronym{mno}{MNO}{Mobile Network Operator}
\newacronym{sla}{SLA}{Service Level Agreement}
\newacronym{embb}{eMBB}{Enhanced Mobile Broadband}
\newacronym{urllc}{uRLLC}{Ultra-Reliable Low-Latency Communications}
\newacronym{v2x}{V2X}{Vehicle-to-Everything}
\newacronym{mmtc}{mMTC}{Massive Machine-Type Communications}
\newacronym{ue}{UE}{User Equipment}
\newacronym{ran}{RAN}{Radio Access Network}
\newacronym{gnb}{gNodeB}{Next Generation Node B}
\newacronym{mitm}{MitM}{Man-in-the-Middle}
\newacronym{nas}{NAS}{Non-Access Stratum}
\newacronym{qoe}{QoE}{Quality of Experience}
\newacronym{qos}{QoS}{Quality of Service}
\newacronym{dos}{DoS}{Denial of Service}
\newacronym{amf}{AMF}{Access and Mobility Management Function}
\newacronym{smf}{SMF}{Session Management Function}
\newacronym{upf}{UPF}{User Plane Function}
\newacronym{ausf}{AUSF}{Authentication Server Function}
\newacronym{nssai}{NSSAI}{Network Slice Selection Assistance Information}
\newacronym{snssai}{S-NSSAI}{Single Network Slice Selection Assistance Information}
\newacronym{aka}{AKA}{Authentication and Key Agreement}
\newacronym{sa}{SA}{Standalone}
\newacronym{nsa}{NSA}{Non-Standalone}
\newacronym{nr}{NR}{New Radio}
\newacronym{epc}{EPC}{Evolved Packet Core}
\newacronym{imsi}{IMSI}{International Mobile Subscriber Identity}
\newacronym{supi}{SUPI}{Subscriber's Permanent Identifier}
\newacronym{suci}{SUCI}{Subscriber Concealed Identifier}
\newacronym{pdu}{PDU}{Protocol Data Unit}
\newacronym{sqn}{SQN}{Sequence Number}
\newacronym{sn}{SN}{Serving Network}
\newacronym{mac}{MAC}{Message Authentication Code}
\newacronym{pfs}{PFS}{Perfect Forward Secrecy}
\newacronym{fbs}{FBS}{Fake Base Station}
\newacronym{cots}{COTS}{Commercial Off-the-Shelf}
\newacronym{dns}{DNS}{Domain Name System}
\newacronym{sdn}{SDN}{Software-defined Networking}
\newacronym{nfv}{NFV}{Network Function Virtualization}
\newacronym{nst}{NST}{Network Slice Template}
\newacronym{sst}{SST}{Slice/Service Type}
\newacronym{sd}{SD}{Slice Differentiator}
\newacronym{mec}{MEC}{Multi-access Edge Computing}
\newacronym{sdr}{SDR}{Software-Defined Radio}
\newacronym{plmn}{PLMN}{Public Land Mobile Network}
\newacronym{autn}{AUTN}{Authentication Token}
\newacronym{abba}{ABBA}{Anti-Bidding down Between Architectures}
\newacronym{kdf}{KDF}{Key Derivation Function}
\newacronym{snn}{SNN}{Serving Network Name}
\newacronym{at}{AT}{Algorithm Type}
\newacronym{aid}{AId}{Algorithm Identifier}
\newacronym{fc}{FC}{Function Code}
\newacronym{htb}{HTB}{Hierarchical Token Bucket}
\newacronym{tbf}{TBF}{Token Bucket Filter}
\newacronym{kpi}{KPI}{Key Performance Indicator}
\newacronym{rtt}{RTT}{Round-Trip Time}
\newacronym{ids}{IDS}{Intrusion Detection System}
\newacronym{nssf}{NSSF}{Network Slice Selection Function}

\journal{Computers & Security}

\newcolumntype{C}[1]{>{\centering\arraybackslash}m{#1}}
\newcolumntype{L}[1]{>{\raggedright\arraybackslash}m{#1}}

\begin{document}

\begin{frontmatter}

\title{Integrity Under Siege: A Rogue gNodeB's Manipulation of 5G Network Slice Allocation} %% Article title

%% Author name
\author[inria]{Jiali Xu}
\author[inria]{Valeria Loscri}
\author[inria,univ-lille]{Romain Rouvoy}

\affiliation[inria]{organization={Inria centre at the University of Lille},
            addressline={40 Avenue Halley}, 
            city={Villeneuve-d'Ascq},
            postcode={59650}, 
            state={Nord},
            country={France}}

\affiliation[univ-lille]{organization={University of Lille, CNRS, UMR CRIStAL},
            addressline={Avenue Henri Poincaré},
            city={Lille},
            postcode={59000},
            state={Nord},
            country={France}}

%% Abstract
\begin{abstract}
The advent of 5G networks, with network slicing as a cornerstone technology, promises customized, high-performance services, but also introduces novel attack surfaces beyond traditional threats. 
This article investigates a critical and underexplored integrity vulnerability: the manipulation of network slice allocation to compromise \emph{Quality of Service} (QoS) and resource integrity.
We introduce a threat model, grounded in a risk analysis of permissible yet insecure configurations like null-ciphering (5G-EA0), demonstrating how a rogue gNodeB acting as a Man-in-the-Middle can exploit protocol weaknesses to forge slice requests and hijack a \emph{User Equipment}'s (UE) connection.
Through a comprehensive experimental evaluation on a 5G testbed, we demonstrate the attack's versatile and severe impacts.
Our findings show this integrity breach can manifest as obvious QoS degradation, such as a 95\% bandwidth reduction and 150\% latency increase when forcing UE to a suboptimal slice, or as stealthy slice manipulation that is indistinguishable from benign network operation and generates no core network errors.
Furthermore, we validate a systemic resource contamination attack where redirecting a crowd of UE orchestrates a Denial-of-Service, causing packet loss to exceed 60\% and inducing measurable CPU saturation ($\approx$80\%) on core network \emph{User Plane Functions} (UPFs).
Based on these results, we discuss the profound implications for \emph{Service Level Agreements} (SLAs) and critical infrastructure.
We propose concrete, cross-layer mitigation strategies for network operators as future work, underscoring the urgent need to secure the integrity of dynamic resource management in 5G networks.
\end{abstract}


\end{frontmatter}


\newcommand{\packet}[1]{\texttt{#1}}
\newcommand{\field}[1]{\textit{#1}}
\newcommand{\key}[1]{$K_{\mathrm{#1}}$}

\section{Introduction}

The fifth generation (5G) of cellular technology represents a paradigm shift from a monolithic network architecture to a dynamic, programmable platform. 
A key enabler of this transformation is network slicing, which allows \glspl{mno} to partition a single physical network infrastructure into multiple, virtual, end-to-end networks~\cite{afolabi2018network}. 
Each slice can be customized with specific characteristics, such as high bandwidth, ultra-low latency, or massive connectivity, to meet the diverse \gls{qos} requirements of different applications. 
These include high-bandwidth \gls{embb} for applications like 4K video streaming and augmented reality; \gls{urllc} for mission-critical services such as remote surgery, industrial automation, and \gls{v2x} communication; and \gls{mmtc} to connect billions of low-power sensors in smart cities and grids~\cite{3gpp2024systemarchitecture}. This capability unlocks new business models and services, but simultaneously introduces complex security challenges.

Although 5G network security has received considerable attention, the focus has primarily been on protecting against threats to confidentiality and authenticity.
Research has extensively covered inherited vulnerabilities from previous generations, the privacy implications of subscriber identifiers~\cite{dabrowski2014imsi}, and the formal security of authentication protocols~\cite{hussain20195greasoner}.
However, there is a critical research gap when considering the integrity of dynamic resource management processes.
The mechanisms that govern how \gls{ue} is allocated to a specific network slice represent a new and potent attack surface. 

The core statement of this work is that the integrity of this slice allocation process is a fragile dependency. 
An adversary who can manipulate this allocation process can directly undermine the core value proposition of 5G, which is the guaranteed delivery of customized \gls{qos}, without necessarily needing to decrypt user data. 
While 5G security standards define robust cryptographic protections for signaling between the \gls{ue} and the core network, they implicitly trust the \gls{ran} element, known as \gls{gnb}, to be a faithful intermediary~\cite{cao2019survey}. Nevertheless, the threat from rogue \glspl{gnb} is a well-established model in the existing literature~\cite{singla2021look,rupprecht2019breaking,mubasshir2024fbsdetector,lotto2023baron}. This oversight is significant, as integrity and availability attacks targeting resource allocation can have severe operational and economic consequences, from violating \glspl{sla} to disrupting critical services.

This paper bridges this gap by providing a comprehensive analysis of an integrity attack on the 5G network slice allocation process, executed by a rogue \gls{gnb}. We move beyond theoretical discussions by grounding our threat model in a risk analysis of standards-compliant but insecure network configurations and the exploitation of documented protocol weaknesses. We then present a multifaceted experimental evaluation that demonstrates the practical feasibility and diverse impacts of this threat.

The main contributions of this work are as follows:
\begin{itemize}
\item \textbf{A Refined Threat Model:}
We detail a practical threat model for slice misallocation by a rogue \gls{gnb}, demonstrating how it can intercept and forge slice requests by exploiting plaintext visibility from null-ciphering configurations and predictable elements in the \gls{nas} registration procedure.

\item \textbf{Comprehensive Experimental Validation:}
We implement our attack in a 5G \gls{sa} testbed and demonstrate its versatile and severe impacts across a range of simulated real-world scenarios. 
We prove the attack can manifest as (a) stealthy slice manipulation by exploiting the protocol vulnerabilities;
(b) severe \gls{qos} degradation, where forcing \gls{ue} from an eMBB to an mMTC slice results in a 95\% bandwidth reduction;
and (c) systemic resource contamination, where redirecting a crowd of \gls{ue} orchestrates a \gls{dos} that causes measurable CPU saturation on core network \glspl{upf}.

\item \textbf{Analysis of Implications and Future Mitigation:}
We analyze the profound implications of these attacks for \glspl{sla} and critical infrastructure.
Based on this analysis, we propose a concrete mitigation framework including core network anomaly detection, cross-layer correlation, and preventative policy hardening, as a direction for future research and implementation.
\end{itemize}

The remainder of this paper is structured as follows. 
Section~\ref{sec:background} provides background on 5G architecture and network slicing and Section~\ref{sec:literature} reviews related work. 
Section~\ref{sec:threatmodel} details our threat model and its exploitable conditions. 
Section~\ref{sec:vulnerabilities} analyzes the specific vulnerabilities in the network slicing procedure that enable the attack. 
Section~\ref{sec:evaluation} presents the setup and results of our comprehensive experiments. 
Section~\ref{sec:discussion} discusses the broader implications of our findings and proposes mitigation strategies. 
Finally, Section~\ref{sec:conclusion} concludes the paper.

\section{Background on 5G Slicing and Security} \label{sec:background}

This section provides the necessary technical background on the 5G \gls{sa} architecture, the network slicing mechanism, and the \gls{nas} protocol that governs the registration procedure targeted by our attack.

\subsection{5G Network Slicing Architecture}

5G networks are designed to support a heterogeneous set of services on a common physical infrastructure. 
Network slicing allows an \gls{mno} to instantiate multiple logical networks, or slices, on this infrastructure, where each slice has its own set of network functions and resource policies. 
Each slice is identified by a \gls{snssai}, which is composed of a mandatory 8-bit \gls{sst} and an optional 24-bit \gls{sd}~\cite{3gpp2023digital}. 
The \gls{sst} defines the slice's primary characteristic. 
For example, the 3GPP specification~\cite{3gpp2024systemarchitecture} outlines several standardized types, including \gls{embb} for high throughput, \gls{urllc} for mission-critical low-latency needs, and \gls{mmtc} for massive IoT connectivity. 
The \gls{sd} further distinguishes slices of the same type. 
\gls{ue} may be subscribed to multiple \glspl{snssai}.

During the initial registration procedure in a 5G \gls{sa} deployment, the \gls{ue} can include a list of preferred slices in its \packet{REGISTRATION REQUEST} message, known as the \field{Requested NSSAI}. 
The \gls{amf} in the core network is responsible for slice selection. 
The \gls{amf} consults the \gls{ue}'s subscription profile, which contains the slices the \gls{ue} is allowed to use, and the capabilities of the serving \gls{gnb}. 
Based on this information, the \gls{amf} partitions the requested slices into an \field{Allowed NSSAI} and potentially a \field{Rejected NSSAI}. 
The allowed slice list is sent back to the \gls{ue} in the \packet{REGISTRATION ACCEPT} message, after which the \gls{ue} establishes a \gls{pdu} session on one of the allowed slices.

\paragraph{\acrlong{qos} and Edge Computing}
A key motivation for network slicing is to meet specific \gls{qos} requirements. 
With the rise of \gls{mec}, 5G networks can place application servers at the network edge to reduce latency. 
Typically, only certain slices are integrated with local \gls{mec} resources. 
For example, an augmented reality application might use a special \gls{urllc} slice that connects the \gls{ue} to a nearby edge cloud. 
If that \gls{ue} was instead placed on a generic \gls{embb} slice without access to the local \gls{mec}, its traffic would have to traverse to a distant data center, incurring higher latency and rendering the application unusable.

\begin{figure}[ht]
\centering
\includegraphics[width=\linewidth]{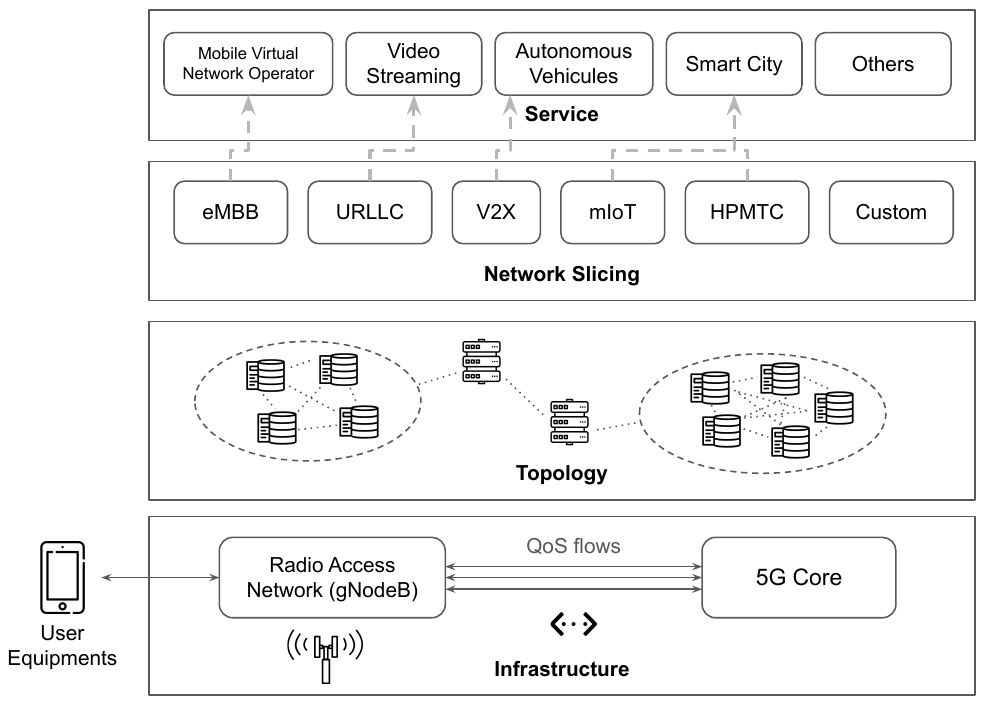}
\caption{5G network slicing architecture.} \label{fig:slicing-arc}
\end{figure}

\subsection{NAS Registration and Security Context Establishment}

Communication between the \gls{ue} and the \gls{amf} is handled by the \gls{nas} protocol. 
\gls{nas} messages are carried over the radio via the \gls{gnb} but are logically end-to-end between the \gls{ue} and \gls{amf}. 
According to 3GPP specifications~\cite{3gpp2024securityarchitecture}, integrity protection for \gls{nas} messages is mandatory, but encryption is optional. 
Operators may choose not to encrypt \gls{nas} messages by using a default "null ciphering algorithm" (5G-EA0), a configuration that has been observed in commercial networks~\cite{jones2025demystifying,chlosta2019lte}, which creates a significant vulnerability.

The \gls{nas} security setup is a multi-step process with a critical design complexity.

\begin{itemize}
\item Initial Unprotected Request:
When \gls{ue} first connects, it sends an initial \packet{REGISTRATION REQUEST} message. 
This message, which can contain the \gls{ue}'s identity (e.g., \gls{suci}) and its \field{Requested NSSAI}, is transmitted in plaintext without any integrity protection or encryption, as no security context yet exists. 
An attacker can intercept this message.   

\item \gls{aka}:
The \gls{amf} receives the unprotected request and initiates the 5G-\gls{aka} procedure to authenticate the \gls{ue} and establish shared secret keys.

\item Security Mode Command:
After successful authentication, the \gls{amf} derives security keys, including \key{NASint} for integrity. 
It then sends a \packet{SECURITY MODE COMMAND} to the \gls{ue}, specifying the chosen cryptographic algorithms (e.g., 5G-IA2 for integrity, and potentially 5G-EA0 for null encryption).

\item Protected Request and Context Activation:
The \gls{ue} uses the command to establish its security context.
To protect the parameters of the initial request from tampering, the \gls{ue}'s \packet{REGISTRATION REQUEST} is re-sent, this time embedded inside the \packet{SECURITY MODE COMPLETE} message.
This message is the first uplink \gls{nas} message to be integrity-protected with the new key, \key{NASint}. 
The \gls{amf} verifies the integrity of this message to make the final slice allocation decision.
\end{itemize}

This design means the slice request appears twice, once unprotected, and again protected.
The 5G architecture implicitly trusts the \gls{gnb} to faithfully relay these messages.
The core network assumes that if the integrity check on the \packet{SECURITY MODE COMPLETE} message passes, the embedded request is authentic and is exactly what the \gls{ue} sent.
This trust, combined with the plaintext visibility afforded by null-ciphering and the complex two-step request process, creates the opportunity for a rogue \gls{gnb} to execute a sophisticated integrity attack.

\section{Related Work} \label{sec:literature}

To contextualize our specific threat model, this section reviews the foundational literature.
We first establish the feasibility of the attacker's posture by examining inherited vulnerabilities from 4G/LTE, known weaknesses in the 5G-AKA protocol, and the well-documented capabilities of rogue base stations. 
We then survey the existing paradigms in network slice security, identifying a critical research gap in the integrity of the initial slice allocation process, which this paper directly addresses.

\subsection{Legacy Vulnerabilities in 5G: The 4G/LTE Inheritance}

The transition to 5G is not a complete replacement of prior infrastructure but an evolution.
The initial and still widespread deployment model for 5G is the \gls{nsa} architecture, which leverages the 5G \gls{nr} standard for the air interface while continuing to utilize the existing 4G \gls{epc} for control plane functions~\cite{fehmi20225g}.
This hybrid configuration means that 5G \gls{nsa} networks are not only subject to new 5G-specific threats but also inherit a wide range of well-recognized vulnerabilities from the 4G/LTE ecosystem~\cite{park20215g}.
This backward compatibility, while essential for ensuring widespread coverage and a smooth migration path, provides a fertile ground for attackers to exploit legacy weaknesses to compromise next-generation services.

\citet{hussain2019privacy} stated in their work that critical flaws persist in foundational protocols, such as the paging mechanism.
This attack vector remains exploitable for user location tracking, spoofed message injection, and \gls{dos} attacks.
Furthermore, long-standing threats like jamming~\cite{arjoune2020smart,flores2023implementation} and the transmission of unsecured pre-authentication traffic~\cite{zhao20215g,bitsikas2021don} continue to affect all mobile network generations, including 5G.
The persistence of implementation-level weaknesses is also a primary concern.
Documented network misconfigurations~\cite{chlosta2019lte} and the continued allowance of weak integrity protection or null encryption schemes~\cite{3gpp2024securityarchitecture,jover2019security,jasim2022analysis} leave 5G deployments susceptible to impersonation attacks and unauthorized data access.

The relevance of these legacy flaws is critically amplified by the susceptibility to downgrade attacks~\cite{karakoc2023never}, which effectively strip 5G devices of their next-generation protections by relegating them to older, more vulnerable networks.
In such attack, a malicious actor forces a 5G-capable device to fall back to a less secure 4G or even 2G network.
This is often achieved by a rogue base station broadcasting a stronger signal for a legacy network type or by manipulating unprotected signaling messages during the pre-authentication phase of the network attachment procedure~\cite{kim20205g}.
Once downgraded, the device is stripped of 5G's advanced security features, such as enhanced encryption protocols and anti-tracking measures, and becomes vulnerable to a host of classic cellular attacks.
These include \gls{imsi} catching, where the \gls{ue}'s permanent identifier is captured in cleartext, and eavesdropping on communications protected by the known-to-be-weak cryptographic algorithms of older standards like GSM~\cite{dabrowski2014imsi}.

Before a secure 5G context is fully established, the continued existence of these legacy systems and their associated vulnerabilities provides a clear and present danger.
The initial interactions between \gls{ue} and the network are a critical window of opportunity for an attacker.
This principle underpins the feasibility of a rogue \gls{gnb} establishing a \gls{mitm} position from which more sophisticated, 5G-specific attacks can be launched.
The security of the entire system is thus constrained by the resilience of its weakest, oldest components, a fact that threat actors consistently exploit~\cite{park20215g}.

\subsection{Formal Analysis of 5G-AKA}

Moving beyond architectural inheritance, the security of 5G networks relies fundamentally on the cryptographic protocols that govern access and key establishment.
The 5G-\gls{aka} protocol is the cornerstone of this security architecture, designed to provide mutual authentication between the \gls{ue} and the network and to establish session keys for protecting user and control plane data~\cite{3gpp2024securityarchitecture}.
While 5G-\gls{aka} introduces notable improvements over its 4G predecessor, including the use of public-key cryptography to conceal the \gls{supi} within a \gls{suci}, it suffers from its own subtle and critical weaknesses.

Rigorous evaluation of the 5G-\gls{aka} protocol using formal analysis methods has provided invaluable insights into its security posture.
\citet{basin2018formal}'s work has revealed that certain security goals are not met under all conditions, particularly in the presence of an active attacker.
These findings expose the foundational cracks that an adversary can exploit to establish a \gls{mitm} position~\cite{xiao2022formal}.
Several key vulnerabilities have been identified from literature:

\paragraph{Lack of Guaranteed Key Confirmation}
A critical finding is that certain essential authentication properties are violated in the period before key confirmation is complete~\cite{basin2018formal}.
The 3GPP standard~\cite{3gpp2024securityarchitecture} does not clearly mandate this final confirmation step in all scenarios, which creates a window where an attacker can manipulate protocol flows before the \gls{ue} and the \gls{sn} have fully established a shared, trusted key.
This pre-authentication vulnerability is a direct enabler for a \gls{mitm} attacker, as it allows for the interception and modification of signaling messages that the \gls{ue} may incorrectly assume are protected~\cite{sultan2025active}.

\paragraph{Traceability and Linkability Attacks}
Despite the introduction of the \gls{suci}, formal analysis has demonstrated that active attackers can still compromise user privacy.
Under certain conditions, an adversary can link a user's \gls{suci} to their long-term identity or track their location and activity across different sessions~\cite{borgaonkar2018new}.
These linkability attacks exploit subtle aspects of the protocol, such as the handling of \glspl{sqn} and re-synchronization error messages, to de-anonymize users~\cite{ko2024formal,cheng2022new}.
This proves that even with enhanced privacy features, the protocol's interactive nature can be leveraged by a sophisticated observer. 

\paragraph{Vulnerability to \gls{mitm} and Impersonation}
The 5G-\gls{aka} protocol, having evolved directly from EPS-\gls{aka}, inherits a susceptibility to \gls{mitm}, impersonation, and replay attacks~\cite{xiao2022formal,xiao20225gaka}.
The complexity of the protocol is a significant contributing factor.
The procedures to handle synchronization failures and \gls{mac} failures create multiple distinct operational cases.
~\citet{xiao2022formal} has shown that an attacker can trigger "cross-attacks" by forcing the protocol from one state to another, exploiting interactions between these cases that would not be apparent from analyzing any single case in isolation.
This highlights a crucial principle: complexity is often the enemy of security.
The very mechanisms designed to make the protocol more robust and resilient to errors can themselves become attack vectors.

\paragraph{Lack of \gls{pfs}} 
A fundamental weakness of the standard 5G-\gls{aka} protocol is its lack of \gls{pfs}.
If an attacker manages to compromise a user's long-term secret key, they can use this key along with previously captured network traffic to compute all past session keys and decrypt historical data~\cite{xiao2022formal,sultan2025active}.
This vulnerability has spurred a significant amount of research into enhanced \gls{aka} protocols that incorporate ephemeral key exchange mechanisms, such as Diffie-Hellman, to ensure that the compromise of a long-term key does not compromise past sessions~\cite{you20235g,ko2024formal}.

The collective findings from the formal analysis of 5G-\gls{aka} are profound.
They demonstrate that the protocol, while strong against passive eavesdroppers, has exploitable logical flaws when subjected to an active \gls{mitm} attacker.
These are not implementation bugs but weaknesses inherent in the standard's specification.
The lack of robust security guarantees in the initial, pre-authentication phases of the protocol provides the precise technical justification for the feasibility of the rogue \gls{gnb}'s \gls{mitm} position, which serves as the foundational premise for the slice allocation integrity attacks detailed later in this work.

\subsection{The Established Threat: Capabilities of a Rogue Base Station}

The concept of a rogue or \gls{fbs} as a threat to cellular networks is well-established and has been a persistent concern through multiple generations of mobile technology~\cite{mubasshir2024fbsdetector}.
An \gls{fbs} is a device, often built from inexpensive \gls{cots} hardware and open-source software, that impersonates a legitimate cell tower.
By broadcasting a signal that appears to be from a legitimate \gls{mno}, often at a higher power level than nearby authentic towers, it can trick \gls{ue} in its vicinity into connecting to it~\cite{bitsikas2021don}.
Once the \gls{fbs} has successfully interposed itself as a \gls{mitm} between the \gls{ue} and the real network, it can launch a variety of attacks targeting the confidentiality, integrity, and availability of the user's communications.

While 5G introduces countermeasures, such as the encryption of the \gls{supi} to prevent classic \gls{imsi} catching, the fundamental threat of the \gls{fbs} remains potent, particularly in hybrid 5G \gls{nsa} environments and by exploiting vulnerabilities in unprotected broadcast signaling~\cite{bitsikas2021don,saedi2022synthetic}.
The demonstrated capabilities of these malicious entities validate the assumed position of the attacker in the threat model under consideration. 

Key attacks that have been extensively documented and demonstrated include:

\paragraph{Downgrade Attacks}
This is one of the most common and effective attacks executed by an \gls{fbs}.
The attacker exploits the backward compatibility of mobile devices by broadcasting a signal for a legacy network technology~\cite{karakoc2023never}.
Because \gls{ue} is designed to connect to the strongest available signal from their carrier, a high-power \gls{fbs} advertising a 4G network can attract the device away from a legitimate 5G signal~\cite{lotto2023baron}.
This forces the \gls{ue} into a security context with known cryptographic weaknesses, enabling the attacker to decrypt calls and data traffic~\cite{dabrowski2014imsi,bitsikas2021don}.
This attack effectively neutralizes many of 5G's security advancements by preventing the \gls{ue} from ever using them.

\paragraph{Signaling and Data Interception}
In its position as a \gls{mitm}, an \gls{fbs} can intercept and modify user data.
A primary vector is \gls{dns} spoofing.
After attaching a \gls{ue} device, the \gls{fbs} can intercept its \gls{dns} queries and return fraudulent IP addresses, redirecting the user to phishing sites or malicious servers designed to harvest credentials or deliver malware~\cite{singla2021look}.
While the widespread adoption of HTTPS provides significant protection by validating server certificates, portions of network traffic remain unencrypted, leaving room for attackers to exploit poorly secured environments or deceive users into disregarding browser security warnings~\cite{bhushan2017man}.

\paragraph{\acrlong{dos}}
An \gls{fbs} can deny service to legitimate users in several ways.
It can simply attach \gls{ue} and then drop all their traffic.
A more sophisticated attack involves exploiting the network attach or handover procedures~\cite{zhao20215g,moheddine2025identifying}.
During the initial attach procedure, before security parameters are fully negotiated, certain signaling messages are sent unprotected.
An \gls{fbs} can send a malicious message, such as an \packet{Attach Reject} with a specific cause code, that instructs the \gls{ue} to no longer attempt to connect to that network or any other LTE/5G network~\cite{singla2021look}.
This can effectively disable the device's cellular connectivity until it is rebooted, turning a radio-level attack into a persistent \gls{dos} condition.

\subsection{Security Paradigms in 5G Network Slicing}

The advent of network slicing in 5G has introduced a new frontier of security challenges.
A comprehensive review of the state-of-the-art by ~\citet{dias20255g} reveals that the academic and industry discourse on network slice security is overwhelmingly dominated by the principle of slice isolation.
This paradigm is foundational to the multi-tenant value proposition of 5G, as it is the primary mechanism that ensures the confidentiality, integrity, and availability are not compromised by the activities or failures of another slice operating on the same shared physical infrastructure~\cite{dhanasekaran2023end}.
The core objective of this research stream is to erect and maintain robust logical barriers between slices, thereby securing the virtualized environment after the slices have been deployed~\cite{olimid20205g}.

The intense focus on isolation is justified by a range of well-defined threats.
The most prominent of these is the cross-slice attack, where an adversary compromises a slice with a lower security posture and uses it as a beachhead to pivot and attack a high-value slice, such as one dedicated to critical infrastructure or enterprise communications~\cite{singh2024security}.
Such attacks typically aim to exploit vulnerabilities in the underlying virtualization layer, including the hypervisor or \gls{sdn} controller.
Another significant concern is resource contention, where a malicious or compromised slice consumes a disproportionate share of physical resources like CPU or spectrum, leading to performance degradation or a \gls{dos} condition for other legitimate slices~\cite{dias20255g}.
Consequently, the prevailing security solutions focus on hardening the virtualization layer, implementing strict access controls, and deploying slice-aware monitoring to detect inter-slice anomalies~\cite{thantharate2020secure5g}.

A secondary, related area of research concentrates on securing the slice management and orchestration plane~\cite{kotulski2018towards,porambage2019secure}.
Research in this domain addresses threats across different stages of the slice lifecycle.
During the preparation phase, studies highlight the risk of an attacker tampering with a \gls{nst}, which could inject malware or misconfigurations into every slice instantiated from that template~\cite{dias20255g,thantharate2020secure5g}.
During the installation and activation phase, the threats involve the compromise of management APIs to create unauthorized slices or maliciously alter a slice's configuration~\cite{olimid20205g}.
To mitigate these risks, researchers advocate for strong authentication and authorization for all management interfaces and the adoption of a Zero Trust Architecture within the core network's control plane~\cite{sathi2018novel,ni2018efficient}.

While these security paradigms are essential, they are based on a critical and largely unexamined assumption: that the initial request for a slice and its subsequent allocation to a user are legitimate and have not been tampered with.
The existing literature focuses extensively on securing the isolation between slices and their management, but overlooks the integrity and the initial signaling that determines how and for whom those slices are constructed.
This reveals a significant gap in the state-of-the-art: the security of the slice allocation procedure itself.
The potential for an attacker to intercept and maliciously modify a user's slice requirements during the initial network registration process is an under-investigated threat vector~\cite{gao2024security}.
This novel threat targets the semantic integrity of the network's high-level procedures, turning resource management capabilities into a weapon.
Therefore, investigating the integrity of the slice allocation process is a critical and necessary step to developing a truly holistic security framework for 5G networks.

\section{Threat Model: A Risk-Based Analysis} \label{sec:threatmodel}

This section details our threat model, which is constructed on a risk-based analysis of permissible network configurations and the exploitation of known protocol weaknesses.

\subsection{Scenario and Attacker Posture}

We consider a deployment of a 5G \gls{sa} network with multiple slices available, such as a high-bandwidth \gls{embb} slice served from a central cloud, an ultra-low-latency slice with local edge processing, and a private slice for a specific enterprise.
In this environment, \gls{ue} is subscribed to these slices, and is configured to actively request specific slices during registration to meet their application needs.
The network consists of standard 5G core functions, including the \gls{amf}, \gls{smf}, and \glspl{upf} associated with different slices, and \gls{gnb} base stations providing radio coverage.

The attacker in our model operates a malicious base station, which we call a rogue \gls{gnb}.
This is a device under the attacker’s control that impersonates a legitimate 5G cell, a classic fake base station setup demonstrated as feasible in prior works~\cite{mubasshir2024fbsdetector,lotto2023baron,bitsikas2021don}.
This impersonation can be achieved using \gls{sdr} tools and open-source \gls{ran} software to broadcast the same \gls{plmn} ID and cell identifiers as the real network, but with a stronger signal to attract nearby \gls{ue} to connect to it.
We assume the rogue \gls{gnb} is within radio range of the victim \gls{ue} and can thus intercept its uplink and downlink communications.

Once victim \gls{ue} attaches to the rogue \gls{gnb}, the attacker’s device acts as a \gls{mitm}, relaying messages between the \gls{ue} and the real core network but with the crucial ability to modify certain messages in transit.
We assume the attacker's goal is not to perform a noisy, easily detectable \gls{dos} by simply dropping packets, but rather to subtly or strategically manipulate the network's resource allocation decisions to achieve objectives such as targeted service degradation, undermining of \glspl{sla}, or disruption of specific network slices.
The specific target of this attack is the \field{Requested NSSAI} field in the \gls{nas} \field{REGISTRATION REQUEST} message and the \gls{snssai} chosen for the \gls{pdu} session.
By altering these fields, the attacker influences which slice the core network allocates to the \gls{ue}.

\begin{figure}[ht]
\centering
\includegraphics[width=\linewidth]{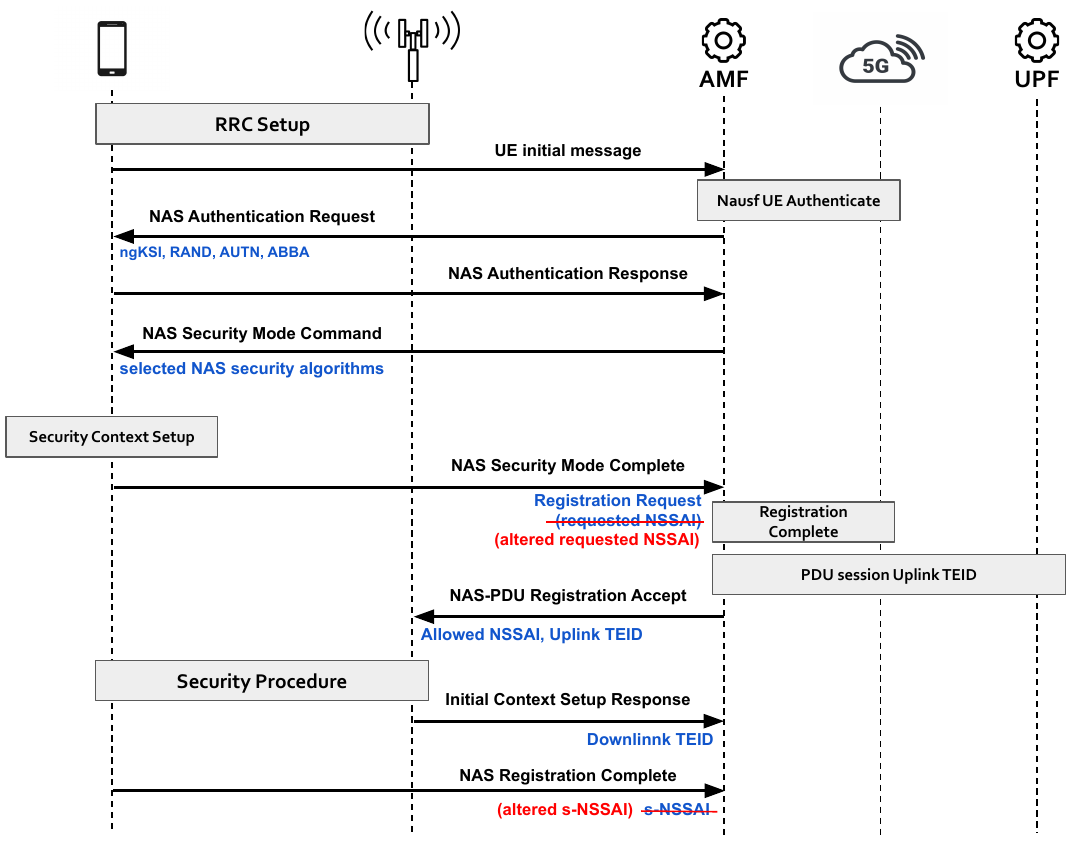}
\caption{The 5G \gls{nas} registration procedure, highlighting relevant message elements (blue) and the rogue \gls{gnb}'s interference points (red).} \label{fig:nas-attack}
\end{figure}

\subsection{Exploitable Conditions}

The success of the attack hinges on two key conditions, which we frame as exploitable vulnerabilities.
These conditions represent a causal chain: the first condition enables the second, creating a clear pathway for the exploit.

\subsubsection{Risk Analysis of Null-Ciphering Configurations}

A cornerstone of our threat model is the exploitation of a standards-compliant but insecure configuration.
The 3GPP specification TS 33.501~\cite{3gpp2024securityarchitecture}, which defines the security architecture for 5G, explicitly permits the use of a "null-ciphering algorithm" identified as 5G-EA0.
When this algorithm is selected, the payload of \gls{nas} messages is not encrypted, although integrity protection may still be applied.

Beyond enabling active manipulation, the use of 5G-EA0 introduces a significant passive threat: the exposure of sensitive slice allocation information.
When null-ciphering is active, the \field{requested NSSAI} within the \gls{nas} messages is transmitted in plaintext, allowing a passive eavesdropper to gather valuable intelligence.
This information is far from benign.
By observing which types of devices request specific slices, an attacker can infer the slice's purpose.
For instance, identifying a slice dedicated to critical industrial IoT or a private enterprise network.
Similarly, the slice requested by \gls{ue} can reveal \gls{ue}'s capabilities and intended use, allowing an adversary to profile potential targets.
This intelligence enables an attacker to perform a cost-benefit analysis, evaluating the value of a target and selecting the most effective attack vector, thereby turning a seemingly simple configuration choice into a critical information leak.

While strong encryption is recommended, network operators may choose to enable 5G-EA0 for specific use cases or network slices.
For example, a slice dedicated to massive IoT deployments with low-power, computationally constrained devices might be configured to use null-ciphering to reduce processing overhead and conserve battery life, under the assumption that the transmitted data is not sensitive.
This operational trade-off between security and efficiency creates a window of opportunity.
The allowance of 5G-EA0 significantly weakens subscriber confidentiality and exposes signaling traffic to both passive eavesdropping and active manipulation by a \gls{mitm} attacker~\cite{jover2019security}.
Our threat model evaluates the security risk posed by such a permissible and existing configuration.

\subsubsection{Protocol Weakness Threatening Integrity Protection}

While 3GPP specifications mandate integrity protection for \gls{nas} messages, the security of this mechanism is critically dependent on the secrecy of the derived \gls{nas} integrity key (\key{NASint}).
A significant protocol weakness exists in the initial registration phase, where a \gls{mitm} attacker can obtain the necessary parameters to independently derive this key, thereby threatening the entire integrity protection scheme.

The vulnerability stems from the fact that several \gls{nas} messages are exchanged in plaintext before a security context is established between the \gls{ue} and the \gls{amf}.
A rogue \gls{gnb} acting as a \gls{mitm} can intercept this initial exchange.
This allows the attacker to eavesdrop on critical security parameters that serve as direct inputs to the key derivation process.

\begin{figure}[ht]
\centering
\begin{tikzpicture}[
    node distance=8mm, 
    box/.style={rectangle,draw,rounded corners,inner sep=6pt,minimum width=20mm},
    lbl/.style={font=\scriptsize,align=left},
    arrlabel/.style={midway,right,font=\scriptsize,align=left}
]
\node[box] (ka) {\key{AUSF}};
\node[box, below=of ka] (ks) {\key{SEAF}};
\node[box, below=of ks] (km) {\key{AMF}};
\node[box, below=of km] (kn) {\key{NASint}};
\node[box, below=of kn] (mac) {$MAC$};
\draw[->] (ka) -- node[midway, right] {$KDF(FC=\text{0x6C}, P_0=\text{SNN}, L_0)$} (ks);
\draw[->] (ks) -- node[midway, right] {$KDF(FC=\text{0x6D}, P_0=\text{IMSI}, L_0, P_1=\text{ABBA}, L_1)$} (km);
\draw[->] (km) -- node[midway, right] {$KDF(FC=\text{0x69}, P_0=\text{AT}, L_0, P_1=\text{AI}, L_1)$} (kn);
\draw[->] (kn) -- node[midway, right] {$NIA(cnt=0, dir=\text{"uplink"}, bearer=\text{"3gpp"})$} (mac);
\end{tikzpicture}
\caption{Key-derivation hierarchy from \key{AUSF} to \key{NASint}, detailing KDF function codes and parameters ($P_n$), where $L_n$ denotes the octet length of $P_n$.}
\label{fig:kdf-hierarchy}
\end{figure}
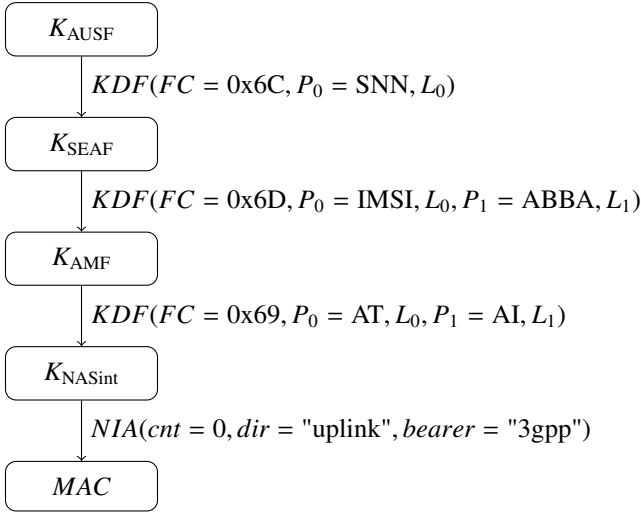

As specified by 3GPP~\cite{3gpp2024securityarchitecture}, the key derivation hierarchy is illustrated in Figure~\ref{fig:kdf-hierarchy}.
It shows how a series of keys are sequentially generated, starting from a high-level key in the \gls{ausf} and culminating in the \gls{nas} integrity key.
Each step in this chain uses a generic \gls{kdf} (typically HMAC-SHA-256) which takes the preceding key and a set of input parameters ($P_n$) as input.
The specific operation at each step is determined by a \gls{fc}, which are predefined values set by the 3GPP specifications.
Therefore, the security of this process relies on the confidentiality of these input parameters.

However, as summarized in Table~\ref{tab:kdf-hierarchy}, many of these critical input parameters are exposed or can be inferred during the unprotected pre-security phase of communication.
Specifically, an attacker can observe or infer the following:

\begin{itemize}
\item \gls{snn}:
This parameter used to derive \key{SEAF} is a public broadcast value.

\item \gls{autn} and \gls{abba}:
These parameters can be transmitted in plaintext within the initial \gls{nas} \packet{AUTHENTICATION REQUEST} messages. 

\item \gls{imsi}:
If not properly concealed as a \gls{suci}.

\item \gls{at} and \gls{aid}:
The \gls{at} is a fixed value (0x02) for integrity protection, and the \gls{aid} is transmitted in the \gls{nas} \packet{SECURITY MODE COMMAND}.
Both are used to derive the final \key{NASint}.
Table~\ref{tab:ai-values} provides the standard values for the \gls{aid}, which specifies the integrity algorithm to be used.
The choice of \gls{aid} is communicated from the \gls{amf} to the \gls{ue} and is observable by the attacker.
\end{itemize}

The 5G-\gls{aka} protocol, while an improvement over its predecessors, is known to be susceptible to various attacks.
By leveraging these known protocol weaknesses in conjunction with the intercepted and inferred parameters detailed in Table~\ref{tab:kdf-hierarchy}, an attacker is positioned to perform the same key derivation calculations as the legitimate network entities.
The exposure of these fundamental parameters during the unprotected phase of the registration protocol provides an adversary with necessary components to compute the \key{NASint} key, fundamentally undermining the foundation of the \gls{nas} integrity protection mechanism.

\begin{table*}[t]
\centering
\caption{Summary of the 5G \gls{kdf} hierarchy and parameter exposure from \key{AUSF} to \key{NASint}.}
\label{tab:kdf-hierarchy}
\renewcommand{\arraystretch}{1.2}
\begin{tabular}{C{.12\textwidth}C{.12\textwidth}C{.12\textwidth}L{.54\textwidth}}
\hline
\textbf{Derived Key} & \textbf{Base Key} & \textbf{\gls{fc}} & \textbf{Parameters and Observability} \\
\hline
\key{AUSF} & CK, IK & 0x6A &
\gls{snn} is often public or can be inferred. \hfill \break
\gls{autn} appears in \gls{nas} \packet{AUTHENTICATION REQUEST} in pre-security. \\ \hline
\key{SEAF} & \key{AUSF} & 0x6C &
\gls{snn} still inferable. Attacker can use this to continue chain. \\ \hline
\key{AMF} & \key{SEAF} & 0x6D &
\gls{imsi} may be protected as \gls{suci} in modern deployments, but legacy/edge cases expose it. \hfill \break
\gls{abba} appears in \gls{nas} \packet{AUTHENTICATION REQUEST} \\ \hline
\key{NASint} & \key{AMF} & 0x69 &
\gls{at} is defined as 0x02 for integrity ({N-NAS-int-alg})~\cite{3gpp2024securityarchitecture}. \hfill \break
\gls{aid} appears in \gls{nas} \packet{Security Mode Command}. \hfill \break
cnt also used by MAC computation is reset to 0. \\ \hline
\end{tabular}
% \vspace{1mm}
\footnotesize{
\textit{Note.} FC is the Function Code defined by 3GPP TS 33.501, Annex A.2. Under the threat model assumptions (e.g., null-ciphering), parameters marked as observable are exposed on the radio interface, enabling the inference of integrity keys.
}
\end{table*}

\begin{table}[t]
\centering
\caption{5G \gls{nas} integrity \acrfull{aid} as specified in 3GPP TS 33.501 \S5.11.1.2.}
\label{tab:ai-values}
\renewcommand{\arraystretch}{1.2}
\begin{tabular}{L{.08\columnwidth}L{.17\columnwidth}L{.6\columnwidth}}
\hline
\textbf{Value} & \textbf{Algorithm} & \textbf{Description} \\ \hline
0000 & NIA0  & Null Integrity Protection algorithm \\
0001 & 128--NIA1 & 128-bit SNOW 3G based algorithm \\
0010 & 128--NIA2 & 128-bit AES-CMAC based algorithm \\
0011 & 128--NIA3 & 128-bit ZUC based algorithm \\
\hline
\end{tabular}
% \vspace{1mm}
\footnotesize{
\textit{Note.} Each algorithm is identified by a 4-bit value. These \gls{aid} values serve as the input parameter ($P_1$) for deriving \key{NASint}.
}
\end{table}

\section{Vulnerabilities in the Slice Allocation Process} \label{sec:vulnerabilities}

The effectiveness of the threat model described above is amplified by several inherent vulnerabilities within the 5G network slicing mechanism.
These weaknesses make the slice allocation manipulation attack particularly potent and difficult to detect.

\subsection{Limited Entropy and Predictability of Slice Identifiers} \label{sec:entropy}

The \gls{nssai}, which uniquely identifies a network slice, has structural properties that facilitate an attack.
Unlike a sophisticated cryptographic hash or a high-entropy unique identifier, the \gls{nssai}'s structure is simplistic and predictable.
The 8-bit \gls{sst} values are standardized and publicly known (e.g., 1 for \gls{embb}, 2 for \gls{urllc}, 3 for \gls{mmtc}).
The 24-bit \gls{sd}, which could provide more entropy, is often optional or unused in many deployments~\cite{enea2021white}.
This constrained and predictable identifier space is trivial for an attacker to navigate.
It allows an adversary to easily infer or even brute-force valid \gls{nssai} combinations for other slices to which the target \gls{ue} might be subscribed, crafting a falsified request that appears legitimate to the \gls{amf}.

\subsection{The NAS Sequence Number Reset Vulnerability}

According to the 3GPP \gls{nas} protocol specification~\cite{3gpp2024nas}, the \field{Uplink NAS Count}, a sequence number used as a nonce to prevent replay attacks and as a parameter in the integrity algorithm, is reset to zero upon the establishment of a new security context during registration.
This predictability simplifies the attacker's task.
For the very first integrity-protected message which the attacker targets, the sequence number is known to be 0.
This removes a variable from the \gls{mac} calculation, making it easier for the attacker to successfully forge a valid \gls{mac} for their manipulated \packet{REGISTRATION REQUEST} message.

\subsection{Lack of Trusted Feedback from the Core Network}

A fundamental architectural vulnerability is the absence of a direct, cryptographically secure feedback channel from the Core Network (specifically the \gls{amf}) to the \gls{ue} to confirm the final slice allocation.
The \gls{ue} sends its request and later receives a \packet{REGISTRATION ACCEPT} message from the \gls{amf}, but both are relayed by the potentially untrusted \gls{gnb}.
The protocol design places implicit trust in the \gls{gnb} to relay this information faithfully, without providing the \gls{ue} with any mechanism, such as a signed attestation from the \gls{amf}, to independently verify the outcome.
This information asymmetry is precisely what the rogue \gls{gnb} exploits, leaving the \gls{ue} unaware that it has been connected to a suboptimal or malicious slice.

\subsection{Exploitable Slice Selection Fallback Mechanism} \label{sec:exploitable}

The 5G slice selection mechanism is designed with a trade-off that prioritizes connectivity over specificity.
If \gls{ue}'s requested slice is unavailable, or if the \gls{ue} provides no preference, the network is designed to silently fall back to a default slice rather than rejecting the connection.
This design choice, intended for operational robustness, creates a significant vulnerability that an attacker can leverage for a stealthy attack.
By intercepting the \packet{REGISTRATION REQUEST} and simply removing the \field{requested NSSAI} list, the rogue \gls{gnb} can force the \gls{amf} to allocate the default slice.

From the \gls{ue}'s perspective, this outcome is indistinguishable from a benign network condition, such as moving to an area where a preferred enterprise slice is not deployed.
The \gls{ue} assumes its preferred slice was unavailable and accepts the default allocation, which becomes a normal and expected outcome.
The attack is thus masked by legitimate protocol behavior.
Conversely, as long as the attacker injects an \gls{nssai} that is part of the \gls{ue}'s subscription, the network will not alert the \gls{ue} that its original preference was overridden.
The protocol lacks a mechanism to ensure the integrity of the slice request itself, leaving the \gls{ue} unable to determine if the slice it was allocated was the result of its own request or an adversary's manipulation.

\section{Experimental Evaluation} \label{sec:evaluation}

Following our analysis of the underlying vulnerabilities, this section provides the practical validation of our threat model.
We first detail the testbed configuration and the distinct network slices created for our experiments.
We then present the results, which demonstrate the attack's feasibility and quantify its versatile impacts, ranging from stealthy slice manipulation and direct \gls{qos} degradation to systemic resource contamination and the resulting core network resource exhaustion.

\subsection{Testbed Configuration}

To validate our threat model and quantify its impact, we constructed an experimental testbed using open-source software components that comply with 3GPP specifications.
The 5G Core network, including the \gls{amf}, \gls{upf}, and other functions, was implemented using Open5GS~\cite{open5gs}.
The \gls{ue} and \gls{gnb} functionalities were simulated using UERANSIM~\cite{ueransim}, a versatile tool that emulates the \gls{ran} and \gls{ue} protocol stacks.

To implement the attack, we modified the UERANSIM \gls{gnb} source code to include a "Rogue" mode.
When activated, this mode enables the \gls{gnb} to intercept the uplink \packet{SECURITY MODE COMPLETE} message from the \gls{ue}, parse the encapsulated \packet{REGISTRATION REQUEST}, replace the \field{requested NSSAI} list with a target \gls{nssai} specified in its configuration, and forward the fraudulent message to the \gls{amf}.

For our experiments, we implemented three distinct network slices to serve as attack targets, each simulating a standard 5G use case and served by a dedicated \gls{upf} to ensure resource isolation.
The specific capacity shaping and \gls{qos} parameters for each slice were configured on the N6 interface of its respective \gls{upf}, as detailed in Table~\ref{tab:slice-capacity}.
The \gls{embb} slice was configured for high-throughput applications with a \SI{1}{Gbit/s} rate limit, using a \gls{htb} and fq\_codel queueing discipline to provide standard WAN latency (\SI{20}{ms} $\pm$ \SI{2}{ms}) under load.
The \gls{urllc} slice was designed for time-critical services, with a \SI{200}{Mbit/s} rate and a netem queueing layer to enforce a strict latency profile of \SI{2}{ms} $\pm$ \SI{0.5}{ms}.
Finally, the \gls{mmtc} slice was configured for low-bandwidth IoT devices, with a \SI{50}{Mbit/s} rate strictly policed by a \gls{tbf} and a higher latency tolerance of \SI{50}{ms}.

\begin{table*}[t]
\centering
\caption{Per-slice capacity shaping on the N6 interface of each UPF.}
\label{tab:slice-capacity}
\small
\renewcommand{\arraystretch}{1.2}
\setlength{\tabcolsep}{4pt}
% \begin{tabular}{l l l S[table-format=4.0] S[table-format=3.0] l l}
\begin{tabular}{L{.05\textwidth}L{.14\textwidth}L{.17\textwidth}L{.09\textwidth}L{.09\textwidth}L{.19\textwidth}L{.16\textwidth}}
\toprule
\textbf{Slice} & \textbf{sNSSAI (SST/SD)} & \textbf{UPF / N6 iface} & {\textbf{Rate}~(Mbit/s)} & {\textbf{Ceil}~(Mbit/s)} & \textbf{Queueing} & \textbf{Latency Model} \\
\midrule
\gls{embb}  & 1 / {0x010203} & \texttt{upf-embb} / \texttt{n6-embb} & 1000 & 1000 & \gls{htb} $\to$ fq\_codel & none \\
\gls{urllc} & 2 / {0x020304} & \texttt{upf-urllc} / \texttt{n6-urllc} & 200  & 200  & \gls{htb} $\to$ netem $\to$ fq\_codel & \SI{2}{ms} $\pm$ \SI{0.5}{ms} \\
\gls{mmtc}  & 3 / {0x030405} & \texttt{upf-mmtc} / \texttt{n6-mmtc} & 50   & 50   & \gls{htb} $\to$ \gls{tbf} & latency \SI{50}{ms} (bucket) \\
\bottomrule
\end{tabular}
% \vspace{1mm}
\footnotesize{
\textit{Note.} \gls{htb} provides hierarchical rate control; fq\_codel keeps queues short under load; netem introduces controlled delay/jitter for \gls{urllc} realism; \gls{tbf} enforces strict rate policing for \gls{mmtc}.
}
\end{table*}

\subsection{Stealthy Network Slice Manipulation} \label{sec:stealthy}

To demonstrate a stealthy attack that exploits the network's benign fallback mechanism, we deployed our rogue \gls{gnb} to intercept the registration procedure of victim \gls{ue}. 

The \gls{ue} was configured to request a specific, non-default \gls{embb} slice identified by SST=1 and SD=0x000001.
The rogue \gls{gnb} was programmed to perform a subtle manipulation: upon intercepting the \packet{REGISTRATION REQUEST}, it erased the SD field from the \field{requested NSSAI} list before forwarding the message to the \gls{amf}.
This manipulation transforms the request from a specific slice to a generic one (SST=1 only). 

The attack was successful and entirely transparent.
The \gls{amf}, receiving a valid request for the \gls{embb} slice type but without a specific differentiator, followed its standard procedure and allocated the network's default \gls{embb} slice (SST=1, SD=0xffffff) to the \gls{ue}.
An analysis of the \gls{amf} logs in~\ref{apx:slice-interception} confirmed that this process generated no warnings or errors, as allocating a default slice is a legitimate and expected behavior.
This outcome validates the vulnerability outlined in Section~\ref{sec:exploitable}, proving that an attacker can force \gls{ue} onto a different, potentially suboptimal slice in a manner that is indistinguishable from benign network operation, thereby evading detection by core network monitoring systems.

\subsection{Quantifying Direct QoS Degradation}

The most direct impact of slice misallocation is a tangible degradation in \gls{qos}, as each slice is provisioned with distinct performance characteristics.
To quantify this, we first established baseline performance profiles for each of the three slices defined in Table~\ref{tab:slice-capacity}.
Under normal conditions, we connected \gls{ue} to each slice and measured \glspl{kpi}, including bitrate, \gls{rtt}, and jitter. 

\begin{figure*}[ht]
\centering
\begin{subfigure}[t]{0.33\textwidth}
    \centering
    \includegraphics[width=\linewidth]{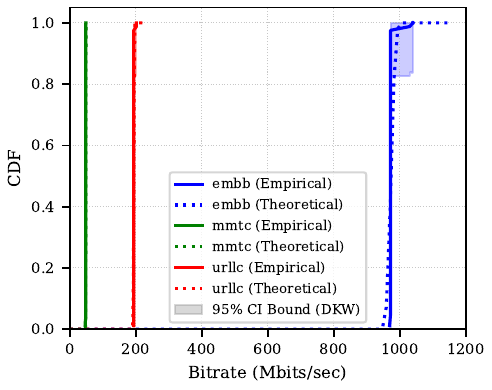}
    \caption{Distribution of Bitrate}
    \label{fig:slice-bitrate}
\end{subfigure}
\hfill
\begin{subfigure}[t]{0.33\textwidth}
    \centering
    \includegraphics[width=\linewidth]{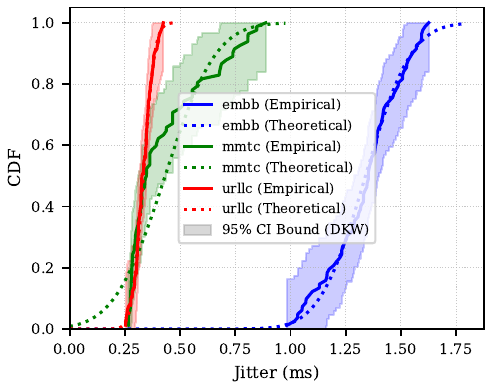}
    \caption{Distribution of Jitter}
    \label{fig:slice-jitter}
\end{subfigure}
\hfill
\begin{subfigure}[t]{0.33\textwidth}
    \centering
    \includegraphics[width=\linewidth]{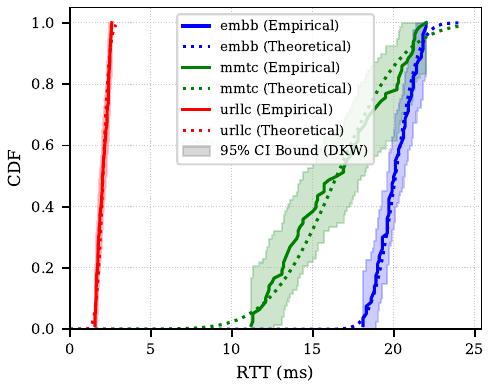}
    \caption{Distribution of RTT}
    \label{fig:slice-rtt}
\end{subfigure}
\caption{Baseline \gls{qos} performance profiles. Empirical and theoretical CDFs for (a) Bitrate, (b) Jitter, and (c) \gls{rtt} across the configured \gls{embb}, \gls{urllc}, and \gls{mmtc} slices.}
\label{fig:slice-qos}
\end{figure*}

The performance differentiation between slices was stark, as visualized in Figure~\ref{fig:slice-qos}.
The bitrate measurements~\ref{fig:slice-bitrate} show the \gls{embb} slice consistently achieving its \SI{1000}{Mbit/s} capacity, while the \gls{urllc} and \gls{mmtc} slices are sharply capped at their respective \SI{200}{Mbit/s} and \SI{50}{Mbit/s} limits.
The \gls{rtt} analysis~\ref{fig:slice-rtt} highlights the ultra-low-latency nature of the \gls{urllc} slice, with 99\% of packets experiencing a latency at or below \SI{2.5}{ms}.
In contrast, the \gls{embb} slice exhibits slightly higher latency, and the \gls{mmtc} slice shows significantly higher latency centered around \SI{50}{ms}.
These results establish a clear performance hierarchy and illustrate the severe consequences of an attack: forcing \gls{ue} that expects \gls{embb} performance onto the \gls{mmtc} slice would result in a 95\% reduction in available bandwidth and a 150\% increase in latency, which is a degradation rendering most broadband applications unusable.

\subsection{Resource Contamination Attack}

Beyond impacting a single user, slice misallocation can be weaponized to create a crowd of devices on a slice not provisioned to handle their traffic, thereby causing resource exhaustion.
To demonstrate this, we simulated an attack where the rogue \gls{gnb} redirects a progressively larger group of \gls{ue}, all running a specific application to a target slice.

\begin{figure*}[ht]
\centering
\begin{subfigure}[t]{0.33\textwidth}
    \centering
    \includegraphics[width=\linewidth]{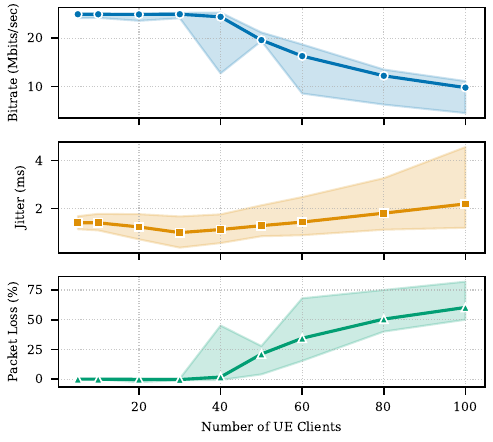}
    \caption{4K Video Streaming on eMBB Slice}
    \label{fig:video-qos}
\end{subfigure}
\hfill
\begin{subfigure}[t]{0.33\textwidth}
    \centering
    \includegraphics[width=\linewidth]{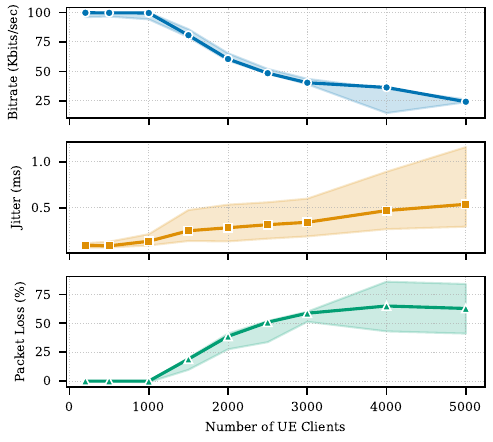}
    \caption{IoT Sensing on URLLC Slice}
    \label{fig:iot-qos}
\end{subfigure}
\hfill
\begin{subfigure}[t]{0.33\textwidth}
    \centering
    \includegraphics[width=\linewidth]{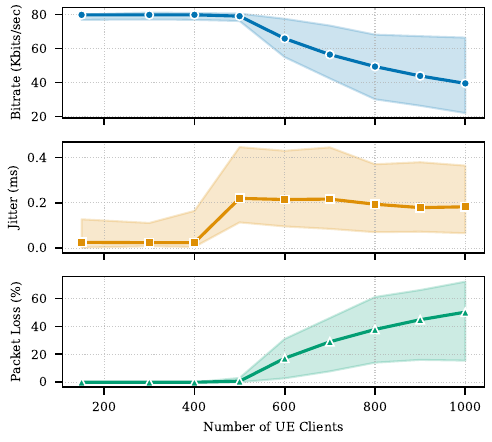}
    \caption{VoIP Telephony on mMTC Slice}
    \label{fig:voip-qos}
\end{subfigure}
\caption{\gls{qos} degradation during the resource contamination attack. Measured Bitrate, Jitter, and Packet Loss as the number of hijacked \gls{ue} clients increases for three scenarios.}
\label{fig:scenario-qos}
\end{figure*}

For example, using the "4K Video Streaming" profile from Table~\ref{tab:traffic-profiles} (requiring \SI{25}{Mbit/s} per user), we hijacked an increasing number of \gls{ue} and forced them onto the targeted \gls{embb} slice.
We measured the received \gls{qos} for each \gls{ue} as the concurrency grew.
The results, shown in Figure~\ref{fig:video-qos}, clearly demonstrate the slice overload.
Up to 40 \gls{ue} devices, the aggregate demand approaches the slice's capacity.
In this range, the bitrate per user remains high (over \SI{20}{Mbit/s}), and packet loss is near 0\%.
However, as more \gls{ue} are forced onto the slice, the total demand exceeds capacity.
This causes the received bitrate for each \gls{ue} to drop sharply, falling to roughly \SI{10}{Mbit/s} with 100 \gls{ue}.
Concurrently, packet loss spikes dramatically (reaching $\approx$60\% at 100 \gls{ue}) and jitter increases as the \gls{upf}'s queues become congested.

We evaluated other scenarios using the same strategy and the results are presented in Figure~\ref{fig:scenario-qos}.
This experiment confirms the attacker's ability to successfully overload a target slice by creating a crowd of redirected \gls{ue}, leading to a self-inflicted \gls{dos} for the entire group of hijacked devices.

We evaluated other scenarios using the same strategy, with results presented in Figure~\ref{fig:scenario-qos}.
For instance, redirecting "IoT Sensing" \gls{ue} (\SI{100}{Kbit/s}) to the \gls{urllc} slice (Figure~\ref{fig:iot-qos}) shows that performance holds until about 1000 clients, after which the bitrate plummets from \SI{100}{Kbit/s} to \SI{25}{Kbit/s} by 5000 \gls{ue}, and packet loss spikes to over 60\%.
Similarly, forcing "VoIP Telephony" \gls{ue} (\SI{80}{Kbit/s}) onto the \gls{mmtc} slice (Figure~\ref{fig:voip-qos}) demonstrates that the slice becomes overloaded after 400 clients, causing the bitrate to drop by 50\% and packet loss to jump to nearly 50\% by 1000 \gls{ue}.
These experiments confirm the attacker's ability to successfully overload any target slice by creating a crowd of redirected \gls{ue}.
This causes a \gls{dos} that impacts not only the group of hijacked devices but also any legitimate \gls{ue} that were properly allocated to the contaminated slice.

\begin{table*}[t]
\centering
\caption{Traffic profiles used to emulate application classes across slices.}
\label{tab:traffic-profiles}
\small
\renewcommand{\arraystretch}{1.2}
\setlength{\tabcolsep}{4.5pt}
\begin{tabular}{L{.18\textwidth}L{.11\textwidth}L{.09\textwidth}L{.1\textwidth}L{.12\textwidth}L{.18\textwidth}L{.1\textwidth}}
\toprule
\textbf{Application} & \textbf{Intended Slice} & \textbf{Type} & {\textbf{Bandwidth}} & {\textbf{Payload}~(Bytes)} & \textbf{Critical KPIs} & \textbf{Concurrency} \\
\midrule
4K Video Streaming & \gls{embb} & UDP CBR & \SI{25}{Mbit/s} & 1470 & Throughput, Packet Loss & 5 - 100 \\
Low-latency IoT Sensing & \gls{urllc} & UDP CBR & \SI{100}{Kbit/s} & 64 & RTT, Packet Loss & 200 - 4000 \\
VoIP Telephony & \gls{mmtc} & UDP CBR & \SI{80}{Kbit/s} & 160 & Jitter, Packet Loss & 150 - 1000 \\
\bottomrule
\end{tabular}

\footnotesize{
\textit{Note.} Concurrency indicates the range of parallel UE clients simulated in the resource contamination experiments (see Figures~\ref{fig:scenario-qos} and~\ref{fig:scenario-perf}).
}
\end{table*}
% & \{5,10,20,30,40,50,60,80,100\} 
% & \{200,500,1000,1500,2000,2500,3000,4000\}
% & \{150,300,400,500,600,700,800,900,1000\} 

\subsection{Core Network Impact}

The resource contamination attack not only degrades the \gls{qos} for the hijacked \gls{ue} but also places a direct and measurable strain on the core network infrastructure.
To quantify this, we monitored the CPU performance of the dedicated \gls{upf} process serving the victim slice during the same experiment described in the previous section.

\begin{figure*}[ht]
\centering
\begin{subfigure}[t]{0.33\textwidth}
    \centering
    \includegraphics[width=\linewidth]{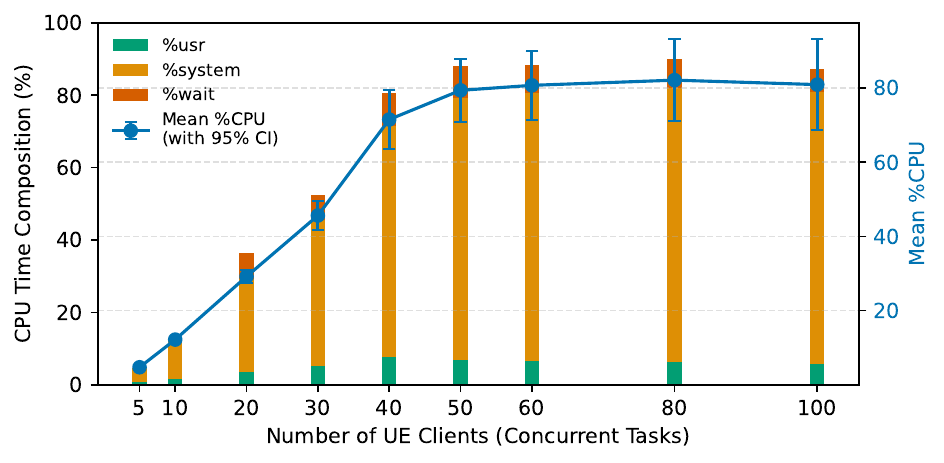}
    \caption{UPF load: 4K Video Streaming on eMBB}
    \label{fig:video-perf}
\end{subfigure}
\hfill
\begin{subfigure}[t]{0.33\textwidth}
    \centering
    \includegraphics[width=\linewidth]{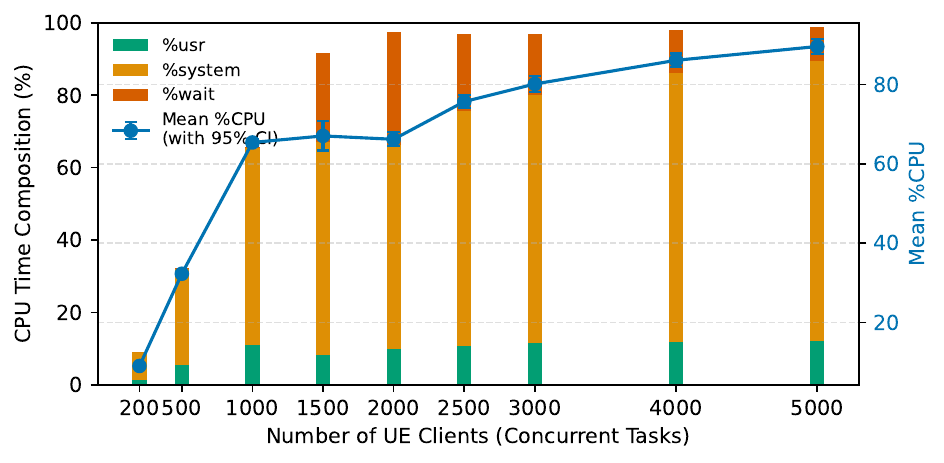}
    \caption{UPF load: IoT Sensing on uRLLC}
    \label{fig:iot-perf}
\end{subfigure}
\hfill
\begin{subfigure}[t]{0.33\textwidth}
    \centering
    \includegraphics[width=\linewidth]{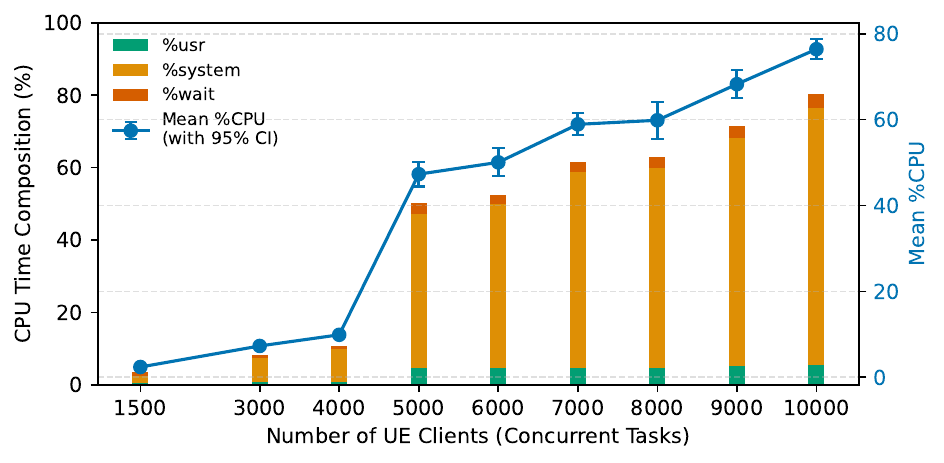}
    \caption{UPF load: VoIP Telephony on mMTC}
    \label{fig:voip-perf}
\end{subfigure}
\caption{Core network impact of the resource contamination attack. \gls{upf} process CPU load (Mean \%CPU) and time composition (stacked bar) as the number of hijacked \gls{ue} clients increases.}
\label{fig:scenario-perf}
\end{figure*}

As the number of concurrent hijacked \gls{ue} was increased, the CPU load on the \gls{upf} scaled in direct correlation with the traffic volume.
Taking the 4K Video Streaming scenario (Figure~\ref{fig:video-perf}) as an example, the \gls{upf}'s \field{'Mean \%CPU'} usage (blue line) climbs steeply, in direct proportion to the number of \gls{ue}, until the slice saturation point of 40 clients.
At this point, the CPU load reaches approximately 80\%. Beyond this saturation point, adding more \gls{ue} (from 40 to 100) causes the CPU load to plateau, indicating it has reached its processing limit.

A deeper analysis of the CPU time composition reveals the nature of this strain.
As the load increases, the time spent in \field{'\%system'} state (orange), handling kernel-level packet processing, increases dramatically, becoming the dominant component of CPU usage.
Crucially, after the 40 \gls{ue} saturation point, the \field{'\%wait'} state (dark orange) begins to appear and grow.
This shift indicates that the \gls{upf} is overwhelmed.
Its CPU is not only busy with kernel tasks but is also increasingly forced to wait for I/O operations, likely as its processing queues and network buffers are congested.

This confirms that the attack successfully translates from a user-level \gls{qos} issue to a tangible resource exhaustion problem for critical core network functions.
The same pattern is visible in the IoT (Figure~\ref{fig:iot-perf}) and VoIP (Figure~\ref{fig:voip-perf}) scenarios, where rising client numbers lead to high CPU utilization dominated by \field{'\%system'} and, eventually, \field{'\%wait'} times.

\section{Discussion} \label{sec:discussion}

This section synthesizes the experimental findings to explore their broader implications for 5G security, discusses the inherent challenges in detecting this novel class of integrity attack, and proposes a multi-layered mitigation strategy. 

\subsection{Implications of Findings}

The experimental validation of the slice allocation attack reveals profound implications that extend beyond simple performance degradation.
The findings demonstrate a fundamental shift in the 5G threat landscape, where the manipulation of network resource management becomes a powerful and stealthy weapon, which poses significant economic, operational, and architectural risks that challenge the foundational security assumptions of 5G.

The experimental results validate our central statement: the integrity of 5G slice allocation is a fragile dependency.
The successful manipulation of a \gls{ue}'s slice assignment has profound implications that challenge 5G's core value proposition, corrupt trust in its service model, and introduce systemic risks to critical infrastructure.

Our findings signal a paradigm shift from traditional threats against confidentiality to attacks targeting the integrity of service.
As our experiments show, this manipulation translates directly into severe \gls{qos} degradation (e.g., a 95\% bandwidth reduction) and core network resource exhaustion.
This proves that the integrity of dynamic resource management is a potent attack surface, subverting 5G's primary promise of guaranteed, differentiated \gls{qos}.

Furthermore, 5G's commercial model relies on network slicing to deliver contractually guaranteed \glspl{sla} for critical applications like remote surgery (\gls{urllc}) or V2X communications.
Our findings show a rogue \gls{gnb} can silently violate these \glspl{sla} by forcing \gls{ue} onto a suboptimal slice.
This transforms a \gls{qos} manipulation into a critical safety and operational risk, fundamentally corrupting trust in 5G's service delivery.
This reveals a dangerous asymmetry: a low-cost attack, feasible with \gls{cots} hardware, can neutralize high-value, high-margin services, posing a significant threat to the 5G enterprise business model.

Finally, as demonstrated by our resource contamination experiment, this vulnerability can be weaponized at scale.
By redirecting a crowd of \gls{ue}, an attacker can orchestrate a \gls{dos} that affects both hijacked and legitimate users, causing tangible resource exhaustion (e.g., \gls{upf} CPU saturation) in the 5G core.
This enables a novel economic denial of service with plausible deniability.
For instance, in a multi-tenant slice, an attacker could use compromised devices from one tenant to flood the resources of a competitor.
The attack originating from a legitimate source would be masked as anomalous resource contention, and transforms the exploit into a sophisticated tool for economic disruption and exposing \glspl{mno} to significant financial and reputational damage from \gls{sla} violations.

\subsection{Challenges in Detection}

This attack is particularly pernicious due to its stealth, which is rooted in its exploitation of legitimate protocol behaviors and fundamental monitoring gaps within the 5G architecture.

The attack's primary camouflage is its reliance on standards-compliant protocol behavior.
As demonstrated in our stealth experiment (Section ~\ref{sec:stealthy}), an attacker can trigger a benign fallback mechanism by simply erasing an optional field from the slice request.
The \gls{amf}, receiving a generic request, correctly follows its standard procedure and allocates the default slice.
This action is considered legitimate, generates no errors in core network logs (~\ref{apx:slice-interception}), and is indistinguishable to the \gls{ue} from benign conditions like network congestion or moving out of a preferred slice's coverage area.
The attack is thus perfectly masked by the network's own mechanisms for operational robustness.

These challenges are compounded by architectural monitoring gaps.
Traditional, single-layer security monitoring is ineffective because of a fundamental partial observability gap.
A monitoring point at the \gls{amf}, for instance, has a critical blind spot: it sees the \gls{gnb}-forwarded \packet{REGISTRATION REQUEST}, which appears valid, but lacks the crucial context of the \gls{ue}'s initial request to detect the discrepancy.
This creates a symptom-cause disconnect, as monitoring the user plane is also insufficient.
An \gls{ids} on the user plane would only observe the consequences of the attack (e.g., low bitrate, high latency), mistaking them for benign network congestion, with no visibility into the control plane manipulation that was the root cause.
The attack strategically creates an inconsistency between protocol steps and network layers, rather than an anomaly within a single layer, rendering conventional detection methods insufficient.

\subsection{Proposed Mitigation Strategies}

Addressing the slice allocation integrity threat requires a defense-in-depth strategy that combines preventative controls to harden the protocol and detective controls to identify attacks that bypass initial defenses.
We propose a three-layer framework targeting the core network, the intersection of network layers, and the \gls{ue} itself.

\subsubsection{Core Network Monitoring and Anomaly Detection}

This detective layer enhances core network intelligence to identify the attack's semantic inconsistencies.
We propose two complementary approaches:

First, an AI/ML-based \gls{ue} behavioral profiling system, potentially co-located with the \gls{amf}, could model each subscriber's (\gls{suci}) typical behavior.
By learning historical slice request patterns and correlating them with contextual data like location (Tracking Area Code) and application type (inferred by the \gls{smf}), the system could flag significant deviations.
For instance, a stationary IoT device suddenly requesting a high-bandwidth \gls{embb} slice would be flagged as a high-risk anomaly, even if its cryptographic check is valid.

Second, as a broader mechanism, the \gls{nssf} could perform real-time, aggregate slice usage analysis.
This system would monitor the statistical distribution of active \gls{ue} across all slices.
An anomalous spike in allocations to a single slice, particularly a default one, coupled with a drop in requests for other slices would strongly indicate a large-scale misallocation or resource contamination attack.

\subsubsection{Cross-Layer Correlation of Control and User Plane}

This strategy introduces a verification loop by correlating control plane directives with user plane performance.
We propose a logical function, a \gls{qos} Integrity Monitor, integrated with the \gls{smf} and \gls{upf}.
During \gls{pdu} session establishment, this monitor would receive the expected \gls{qos} profile (e.g., guaranteed bitrate, max \gls{rtt}) associated with the \gls{ue}'s \field{Allowed NSSAI} from the control plane.
Once the session is active, the monitor would continuously measure the actual user plane \glspl{kpi} (throughput, latency, jitter) at the \gls{upf}.
A persistent, significant mismatch between the expected and measured performance, for instance, measuring \SI{50}{ms} \gls{rtt} for a session that was allocated to a \gls{urllc} slice, would serve as a high-confidence indicator of an integrity breach.
This cross-layer check turns the attack's primary impact of \gls{qos} degradation into the primary signal for its detection.

\subsubsection{Preventative Hardening of Protocols and Policies}

Preventative controls are the most robust solution.
These strategies focus on hardening the protocol and network policies to eliminate the root vulnerabilities.

First, a strict policy mandating the disablement of null-ciphering (5G-EA0) is the most direct countermeasure.
While 5G-EA0 is standards-compliant, it creates the vulnerability window detailed in our threat model.
Enforcing encryption for all \gls{nas} messages would blind the rogue \gls{gnb}, preventing it from reading the initial \field{Requested NSSAI} and disrupting this specific attack chain.

Second, granular slice access control can be enforced as a \gls{ran}-side preventative.
\glspl{mno} should configure each \gls{gnb} with an explicit allow-list of \glspl{snssai} that it is authorized to serve, based on its location and purpose (e.g., a generic \gls{gnb} should not be able to access a sensitive, high-end slice).
This applies a principle of least privilege, containing the blast radius of a rogue \gls{gnb}.

Third, the predictability of \gls{sd} could be reduced. As noted in our vulnerability analysis (Section ~\ref{sec:entropy}), the low entropy of the 24-bit \gls{sd} makes it trivial for an attacker to guess valid identifiers.
We propose that operators use high-entropy values for the \gls{sd}, such as a cryptographic hash of a non-public name, rather than sequential or predictable numbers.
This makes it computationally infeasible for an adversary to infer a slice's purpose or target a specific high-value slice.

\section{Conclusion} \label{sec:conclusion}

This paper has demonstrated that the integrity of the 5G network slice allocation process is a critical, practical, and previously underestimated security concern.
Our primary contribution is a refined threat model, grounded in a risk analysis of standards-compliant but insecure configurations like null-ciphering, which details a plausible pathway for a rogue \gls{gnb} to manipulate a \gls{ue}'s slice assignment.

We validated this threat through a comprehensive experimental evaluation on a 5G testbed, revealing the attack's versatility and severe impacts.
Our results proved the manipulation can manifest as:

\begin{itemize}
\item Obvious \gls{qos} degradation, such as a 95\% bandwidth reduction by forcing \gls{ue} to a suboptimal slice.

\item Stealthy performance bottlenecks that mimic benign network fallback mechanisms, evading simple detection.

\item Systemic resource contamination, where a crowd of hijacked \gls{ue} weaponizes the vulnerability to induce a \gls{dos} and cause tangible CPU exhaustion in core network functions.
\end{itemize}

These findings underscore a fundamental shift in the 5G attack surface, from static data protection to the manipulation of dynamic resource management processes.
This threat directly subverts 5G's core value proposition of guaranteed \gls{qos} and corrupts trust in \glspl{sla}.

Finally, these findings necessitate a move beyond traditional, confidentiality-focused security models.
As a direction for future research, we proposed a concrete, multi-layered mitigation framework.
This strategy combines preventative controls (e.g., disabling 5G-EA0) with detective measures (e.g., core network anomaly detection and cross-layer \gls{qos} integrity monitoring).
Investigating the practical efficacy of these proposed defenses remains an essential next step toward ensuring that the promise of customized, reliable 5G services can be delivered securely.

\section*{Data and Code Availability}
The source code, deployment configurations, and data to reproduce this study are publicly available.
The modified UERANSIM source code, which implements the rogue gNodeB mode, is available at \href{https://github.com/jxu96/UERANSIM}{github.com/jxu96/UERANSIM}.
The Open5GS deployment scripts and Docker-based configurations used to instantiate the 5G core network and distinct slices are available at \href{https://github.com/jxu96/Open5GS-Docker}{github.com/jxu96/Open5GS-Docker}.
The raw and processed datasets generated and analyzed during the QoS and resource contamination experiments  are available within these repositories.

\section*{Disclosure of Funding Sources}
This research was made possible with support from the Horizon Europe research and innovation program of the European Union under grant agreement number 101092912 (project MLSysOps) and COST Action CA 22104 BEiNG-WISE.

\appendix
\lstdefinelanguage{Syslog}{
  morekeywords={INFO,WARN,WARNING,ERROR,DEBUG,CRITICAL,NOTICE},
  sensitive=true,
}

\lstdefinestyle{logstyle}{
  language=Syslog,
  basicstyle=\ttfamily\footnotesize,
  keywordstyle=\bfseries,
  commentstyle=\itshape\color{OliveGreen},
  numbers=left, numberstyle=\tiny\color{black}, numbersep=8pt,
  breaklines=true, breakatwhitespace=true, columns=fullflexible,
  keepspaces=true, showstringspaces=false, tabsize=2,
  frame=single, rulecolor=\color{gray}, framesep=3pt,
  captionpos=b,
  aboveskip=6pt, belowskip=6pt,
  postbreak=\mbox{\textcolor{gray}{\(\hookrightarrow\)}\space},
}

\section{Slice Allocation Interception} \label{apx:slice-interception}

Listing~\ref{lst:amf-logs} provides a detailed log excerpt from the Access and Mobility Management Function (AMF) during the stealthy slice manipulation attack described in Section~\ref{sec:stealthy}. These logs validate the attack's stealthy nature and its ability to evade detection by the core network, as no errors are generated.

The key event is captured in Line 15. Here, the AMF log confirms the allocation of the S-NSSAI [SST:1 SD: 0xffffff] to the User Equipment (UE). This is the network's default eMBB slice, not the specific slice $(SST=1, SD=0\times000001)$ that the UE originally requested. The rogue gNodeB successfully manipulated the registration request by erasing the SD field, forcing this benign fallback.

Critically, the log shows the registration completes successfully (Line 10) and the entire procedure is logged as normal INFO operation. This lack of errors confirms that the attack is indistinguishable from a legitimate network condition (e.g., a UE requesting a generic slice), effectively masking the manipulation.

\begin{lstlisting}[style=logstyle,
  caption={AMF Logs}, label={lst:amf-logs}]
10/23 20:11:35.245: [amf] INFO: InitialUEMessage (../src/amf/ngap-handler.c:401)
10/23 20:11:35.245: [amf] INFO: [Added] Number of gNB-UEs is now 1 (../src/amf/context.c:2550)
10/23 20:11:35.245: [amf] INFO:     RAN_UE_NGAP_ID[1] AMF_UE_NGAP_ID[1] TAC[1] CellID[0x10] (../src/amf/ngap-handler.c:562)
10/23 20:11:35.245: [amf] INFO: [suci-0-001-01-0000-0-0-0000000001] Unknown UE by SUCI (../src/amf/context.c:1835)
10/23 20:11:35.245: [amf] INFO: [Added] Number of AMF-UEs is now 1 (../src/amf/context.c:1616)
10/23 20:11:35.245: [gmm] INFO: Registration request (../src/amf/gmm-sm.c:1165)
10/23 20:11:35.245: [gmm] INFO: [suci-0-001-01-0000-0-0-0000000001]    SUCI (../src/amf/gmm-handler.c:166)
10/23 20:11:35.279: [sbi] INFO: [UDM] (SCP-discover) NF registered [cc1871ca-aff6-41f0-8671-1b208ef849f0:1] (../lib/sbi/path.c:211)
10/23 20:11:35.284: [sbi] WARNING: [UDM] (SCP-discover) NF has already been added [cc1871ca-aff6-41f0-8671-1b208ef849f0:2] (../lib/sbi/path.c:216)
10/23 20:11:35.504: [gmm] INFO: [imsi-001010000000001] Registration complete (../src/amf/gmm-sm.c:2146)
10/23 20:11:35.504: [amf] INFO: [imsi-001010000000001] Configuration update command (../src/amf/nas-path.c:612)
10/23 20:11:35.504: [gmm] INFO:     UTC [2025-10-23T18:11:35] Timezone[0]/DST[0] (../src/amf/gmm-build.c:559)
10/23 20:11:35.504: [gmm] INFO:     LOCAL [2025-10-23T20:11:35] Timezone[7200]/DST[1] (../src/amf/gmm-build.c:564)
10/23 20:11:35.506: [amf] INFO: [Added] Number of AMF-Sessions is now 1 (../src/amf/context.c:2571)
10/23 20:11:35.506: [gmm] INFO: UE SUPI[imsi-001010000000001] DNN[internet] S_NSSAI[SST:1 SD:0xffffff] smContextRef [NULL] (../src/amf/gmm-handler.c:1241)
10/23 20:11:35.506: [gmm] INFO: SMF Instance [cc3cadba-aff6-41f0-a8af-1920e30e389d] (../src/amf/gmm-handler.c:1280)
10/23 20:11:35.553: [amf] INFO: [imsi-001010000000001:1:11][0:0:NULL] /nsmf-pdusession/v1/sm-contexts/{smContextRef}/modify (../src/amf/nsmf-handler.c:837)
\end{lstlisting}

% \lstinputlisting[style=logstyle, firstline=120, lastline=165,
%   caption={UE Configuration}, label={lst:ue-conf}]{listings/open5gs-ue.yaml}

\bibliographystyle{elsarticle-num-names}
\bibliography{src/references}
\end{document}